\documentclass
[superscriptaddress,secnumarabic,amssymb,amsmath,nobibnotes,aps,prd,showpacs,nofootinbib]{revtex4}%
\usepackage{graphicx,subcaption}
\usepackage{dcolumn}
\usepackage{bm}
\usepackage{xcolor}
\usepackage{amsmath}
\usepackage{amsfonts}
\usepackage{amssymb}
\usepackage{comment}
\usepackage{enumitem}
\usepackage{float}
\usepackage{hyperref}
\usepackage{footnote}

\newcommand{\ed}{\end{document}}
\newcommand{\beq}{\begin{equation}}
\newcommand{\eeq}{\end{equation}}
\newcommand{\beqa}{\begin{eqnarray}}
\newcommand{\eeqa}{\end{eqnarray}}
\newcommand{\bc}{\begin{center}}
\newcommand{\ec}{\end{center}}

\newcommand{\ba}{\begin{array}}
\newcommand{\ea}{\end{array}}

\begin{document}


\title{Explanation of the Generalizations of   Uncertainty Principle from \\
Coordinate and Momentum Space Periodicity}

\author{Subir Ghosh}
\email{subirghosh20@gmail.com}
\affiliation{Physics and Applied Mathematics Unit, Indian Statistical Institute, \\203 B. T. Road, Kolkata 700108, India}

\begin{abstract}
\noindent	
{\bf{Abstract:}} Generalizations of coordinate $x$-momentum $p_x$ Uncertainty Principle, with $\Delta x$ and $\Delta p_x$ dependent terms ($\Delta$ denoting standard deviation), $$\Delta x \Delta p_x\geq i\hbar (1+\alpha\Delta p_x^2 +\beta \Delta x^2)$$ have provided rich dividends as a poor person's approach towards Quantum Gravity, because these can introduce coordinate and momentum scales ($\alpha,\beta$ ) that are appealing conceptually. However, these extensions of Uncertainty Principle are purely phenomenological in nature. Apart from the inherent ambiguity in their explicit structures, the introduction of generalized commutations relations compatible with the the uncertainty relations has some drawbacks. 

In the present paper we reveal that these generalized Uncertainty Principles can appear in a perfectly natural way, in canonical quantum mechanics, if one assumes a periodic nature in coordinate or momentum space, as the case may be. We bring in to light quite old, (but not so well known), works by Judge and by Judge and Lewis, that explain in detail how a consistent and generalized Uncertainty Principle is induced in the case of angle $\phi$ - angular momentum $L_z$, $$\Delta \phi \Delta L_z \geq   i\hbar (1 +\nu \Delta \phi^2)$$ purely from a consistent implementation of {\it{periodic}} nature of the angle variable $\phi $, without changing the $\phi, L_z$ canonical commutation relation. {\it{Structurally this is identical to the well known Extended Uncertainty Principle.}} We directly apply this formalism  to formulate the  $\Delta x \Delta p_x $ Extended Uncertainty Principle. We identify $\beta$ with an observed length scale relevant in astrophysics context. We speculate about the $\alpha$ extension.

\end{abstract}

 
\maketitle



{\bf{Introduction:}} Absence of a fundamental theory of Quantum Gravity has given rise to phenomenological models of generalized uncertainty relations, that in a top-down approach, tries to incorporate signatures of Quantum Gravity. Most notable of such effects is the proposed existence of a minimum length scale, adhering to String theory ideas. This can be achieved by considering, (contrary to the Heisenberg Uncertainty Principle (HUP) $\Delta x^i\Delta p_j \ge  (\hbar /2)\delta^i_j$), a Generalized Uncertainty Principle (GUP) of a generic form
\begin{equation}
\Delta x^i\Delta p_j \ge  \frac{\hbar}{2}\delta^i_j (1+\alpha\frac{2G}{\hbar c^3}(\Delta p_j)^2)
    \label{1}
\end{equation}
where $\alpha $ is a dimensionless constant \cite{gup}. Minimization of (\ref{1}) for $\Delta x^i$ yields a minimum length of the order of $L_P={\sqrt{\hbar G/c^3}}$, the Planck length. However, this is the minimal form and there exist more elaborate forms of GUP. The non-uniqueness of the GUP structure stems from the absence of a fundamental theory leading to the GUP in a bottom-up approach (for reviews see \cite{rev}). 

The symmetry between $x_i$ and $p_j$ in the canonical commutation relations, $[x_i,p_j]=i\hbar \delta_{ij},~[x_i,x_j]=[p_i,p_j]=0$ suggests the possibility of a complimentary type of UP, popularly known as Extended Uncertainty Principle (EUP), 
\begin{equation}
\Delta x^i\Delta p_j \ge  \frac{\hbar}{2}\delta^i_j (1+\beta\Lambda(\Delta x^i)^2)
    \label{2}
\end{equation}
where $\beta $ is a dimensionless constant and $\Lambda$ is the Cosmological Constant \cite{eup,park}. \footnote{The works \cite{eup,park} discuss the significance of EUP in the context of dS/AdS Black Hole thermodynamics in deriving the Hawking temperature. In this context it is worth mentioning, (although this is not directly related to our work), that in \cite{scar} the author has shown that if some of the weak field results of a particle moving in an effective potential are allowed to be extrapolated up to horizon, a strong field regime, the Hawking temperature for dS/AdS Black Hole can be derived using the conventional Heisenberg form of Uncertainty Principle.  This implies a minimum scale for de Sitter momentum of the order $m_{dS}c=\hbar {\sqrt{\Lambda /3}}$.\footnote{ At the outset, it needs to be pointed out in favour of GUP that there are compelling, albeit heuristic, arguments suggesting some form of momentum dependent GUP-like corrections in the Uncertainty Principle that leads to a minimum length scale; consistency with high energy string scattering amplitude, removing a paradoxical scenario in Black Hole physics which states that unbounded localization of an event can lead to high enough energy density that allow Black Hole formation at that event itself, making it unobservable, ... among others. To the best of our knowledge, so far such intuitive arguments in favour of EUP is probably unavailable although in the present paper, we have managed to provide possible observational evidences. }}
Finally there is  the maximally extended for GEUP comprising of both $(\Delta x^i)^2$ and $(\Delta p_j)^2$ in the RHS \cite{eup} but we will not go into this in the present work. In the present paper we will be concerned mostly with EUP (\ref{2}). 
 
It needs to be stressed that these generalizations of uncertainty relations are essentially phenomenological in nature. The idea of a minimum length scale is appealing from different perspectives; high energy scattering of strings leading to a fundamental length, removing conceptual  paradoxes in Black Hole Physics, ... among others. In order to have a deeper insight into the origin of such non-canonical UPs, a well-studied proposal \cite{kem} has been to replace  the  canonical commutation relations to   generalized set of commutations relations, consistent with GUP, EUP, GEUP. Recall that the canonical (or Heisenberg UP) is compatible with the canonical commutation relations \cite{com}. However, in the present instances, these modified commutators necessary involve operator-valued terms in the RHS, as for example 
\begin{equation}
[ x^i, p_j~]=i\hbar \delta^i_j (1+\alpha\frac{2G}{\hbar c^3}( p)^2)
    \label{4}
\end{equation}
corresponding to the GUP (\ref{1}).  Apart from the technical problems involved in quantizing a theory with operator-valued commutation relations, (although a possible solution is to map non-canonical degrees of freedom to canonical ones via a Darboux-like transformation), it has been observed recently \cite{la} that such generalizations in UPs can give rise to reference-frame dependent minimum length scales and modified commutators can induce equivalence principle violations \cite{sg1}. However, careful later analysis \cite{s1,s2} seem to suggest that purely quantum phenomena, and not entirely GUP, can be responsible for possible equivalence principle violation since quantum deformed commutators do not necessarily correspond to a deformation in classical Poisson brackets.  To overcome the above mentioned issues and simultaneously generate the GUP, EUP or EGUP, it has been suggested in \cite{la} to replace the usual position and momentum measurement operators by appropriate Positive Operator valued Measure, with an inbuilt finite accuracy. The main point of this construction is that one can still use the canonical variables, commutation relations and Hamiltonian with Heisenberg and Schrodinger  equations unchanged. In \cite{rob}, the authors suggest a General Relativistic framework where an effective GUP can be generated.

The take home message of the above discussion: try to derive the generalized UPs keeping the canonical structure (of commutators, Hamiltonian, ..) intact. We resurrect a quite ancient paper by D. Judge \cite{j} to show that it is indeed possible and in the process reveal an oft studied feature of the Universe - it has a periodicity.\\

\vspace{.3cm}
{\bf{Deriving the Extended Uncertainty Principle:}} The work of Judge \cite{j} deals with the direct evaluation of  uncertainty relation involving $z$-component of orbital angular momentum, $L_z=-i\hbar \partial /\partial\phi$ and the angle $\phi$. Naively, it should follow the $p_x,x$ result, that is $\Delta L_z\Delta \phi\ge \hbar/2$ but obviously this is not correct because $\phi$ being an angle, $\Delta \phi \le \pi/{\sqrt{3}}$, corresponding to a uniform distribution $f(\phi )$ of $\phi$ as derived below:
\begin{equation}
for ~ \pi\ge\phi\ge \-\pi, ~ f(\phi)=1/(2\pi), (\Delta\phi )^2=\int _{-\pi}^{\pi}\phi^2 f(\phi)d\phi -(\int _{-\pi}^{\pi}\phi f(\phi)d\phi)^2=\pi^2/3 .
    \label{5}
\end{equation}
Thus as $\Delta L_z $ moves towards zero $\Delta\phi $ has to obey the bound and a contradiction arises. A careful consideration of the fact that wave functions are periodic in $\phi$, Judge \cite{j} was able to derive a modified uncertainty principle 
\begin{equation}
\Delta L_z\Delta \phi\ge \eta\hbar (1-\frac{3(\Delta\phi )^2}{\pi ^2})
    \label{6}
\end{equation}
where $\eta \approx 0.15$ as derived in \cite{j}. Notice that it is structurally identical to the EUP (\ref{2}). Since the work of Judge \cite{ j} is not very well-known, we will repeat some steps of the calculation at the end in an Appendix in our context of EUP. Here we emphasize the fact that the periodic nature of wave  functions of angle variables is  the key factor in deriving the modified UP (\ref{6}); the phase space symplectic structure and associated quantum commutators are strictly canonical.

Substituting $p_x,x$ in place of $L_z, \phi$ we derive (See Appendix for details) the EUP$_p$ (the subscript "p" is to stress that this form is derived here considering only periodic nature of space)  
\begin{equation}
\Delta x\Delta p_x \ge  \frac{\hbar}{2}(1-\frac{12(\Delta x^i)^2}{L^2})
    \label{7}
\end{equation}
where we have assumed the periodicity of wave function $\psi(x=0)=\psi(x=L)$. With $L=L_P/l$ our EUP reduces to
\begin{equation}
\Delta x\Delta p_x \ge  \frac{\hbar}{2}(1-\frac{12l^2(\Delta x^i)^2}{L_P^2}).
    \label{8}
\end{equation}
Before providing justification for the periodic space assumption let us quickly see the immediate consequences of this EUP$_p$. In \cite{park}, generic phenomenological forms of  EUPs,  
\begin{equation} 
\Delta x\Delta p_x \ge  \frac{\hbar}{2}(1\pm\frac{\beta^2(\Delta x^i)^2}{l^2})
    \label{9}
\end{equation}
with $l$ a characteristic length scale and $\beta$ a dimensionless parameter, have been considered. From \cite{park}, it is inferred that the positive sign EUP in (\ref{9}) implies an AdS space, with a lower bound of $\Delta p_x$,  whereas the positive sign EUP in (\ref{9}) can correspond to a dS space. \\
(i) The EUP$_p$ (\ref{7}) is an exact and unambiguous result without any arbitrary parameter since, as we show later, the periodicity parameter $L$ can be identified with robust observational value. Clearly it is  not an outcome of phenomenological and heuristic arguments.\\
(ii) The EUP$_p$ (\ref{7}) leads to an absolute maximum value of $\Delta x\le L/{\sqrt{12}}$ so that $\Delta p_x$ remains positive.\\
(iii) It is interesting to note that, as pointed out in \cite{park}, the maximum value of $\Delta x$ does not have $\hbar$ since it is a consequence of a periodic feature of classical cosmology.

It is interesting to note that, exploiting a possible periodic nature of spacetime, proposed in a "world crystal" model by Kleinert \cite{k}, Jizba, Kleinert and Scardigli \cite{jks} have been able to derive a GUP structure and associated modified $[x,p]$ commutation relations on a spacetime lattice.

{\it{Justification of the spatial periodicity }}: Recent astrophysical observations  \cite{p1}, (corroborating the earlier work \cite{p2}), have shown that, (contrary to the expectation of  galaxies being randomly scattered in the Universe),  mapped in two or three dimensions, galaxies are   clustered on small scales $(\sim 5 h^{-1} Mpc,~h\approx 0.5 - 1)$. In fact, more interesting is the finding as reported in \cite{p2} that galaxies tend to remain correlated over a much larger distance, with a characteristic scale of $\sim 128 h^{-1} Mpc$. The excess of correlation and this apparent periodic distribution of galaxies  suggest possibility of oscillating behaviours in  cosmological evolution. This is not explainable in conventional cosmological models and requires scalar field with non-minimal couplings. In \cite{p3} a model is proposed involving a stable condensate of a pseudo-Goldstone mode having a  periodic distribution of energy density in radial coordinate. \\

\vspace{.3cm}
{\bf{Conclusion:}} Conventionally generalizations of $\Delta x,\Delta p_x$ uncertainty relation is introduced in a purely ad hoc and phenomenological way leading to non-uniqueness in its explicit structure and its subsequent offshoot of non-canonical operator dependent commutation relations yielding further issues. In this perspective, we fall back to the canonical framework and following closely early works by Judge \cite{j} show that well defined Extended Uncertainty Relations can be induced by a periodicity in the space coordinates. We further show that there are observational results showing such periodicity in astrophysical scenario. 

In the present work we used the coordinate representation with $p_x=-i\hbar \partial/\partial x$ and recovered $(\Delta x)^2$ corrections in 
$\Delta x,\Delta p_x$ uncertainty relation. It will be straightforward to consider the complimentary (Fourier space) scenario where $x=i\hbar \partial/\partial p_x$ and generate the GUP with $(\Delta p)^2$ corrections in $\Delta x,\Delta p_x$ uncertainty relation. In this context it might be worthwhile to relate the present work with the earlier idea proposed in \cite{jks}.\\

\vspace{.3cm}
{\bf{Acknowledgement:}} It is a pleasure to thank the anonymous Reviewer for raising many important points and suggestions that have improved the presentation of our work. \\

\vspace{.3cm}
{\bf{Appendix :}} We essentially follow the analysis of Judge \cite{j,j1,j2}, appropriately modified for our system. We assume the periodicity in wave function 
$\psi(x=-L/2)=\psi(x=L/2)$. Define the function
\begin{equation}
V(\gamma )=\int_{-L/2}^{L/2} \psi^*(x+\gamma)x^2\psi(x+\gamma)dx .
    \label{a1}
\end{equation}
Note that $\psi^*(x)\psi(x)$ is identified with an appropriate distribution function. 
The value of $\gamma$, for $L/2\ge \gamma\ge -L/2$, that minimises $V(\gamma )$ (\ref{a1}) is the mean value of $x$ and $(\Delta x)^2$ is the minimum value of $V(\gamma )$. Note that in general this is different from the conventional definition of variances
\begin{equation}
(\Delta x)^2=<(<x-<x>)^2>=<x^2>-<x>^2, (\Delta p_x)^2=<(<p_x-<p_x>)^2>=<p_x^2>-<p_x>^2
    \label{a2}
\end{equation}
where $<A>$ stands for the expectation value of $A$. Indeed, it is easy to check that $(\Delta x)^2$ computed from (\ref{a1}) and (\ref{a2}) are identical for $\infty\ge x\ge -\infty$, that is when the domain of $x$ is the real line and $\psi(\pm \infty)=0$. However for angular variables $\phi$ \cite{j,j1,j2} or for other  periodic systems, ( as the model proposed here), the conventional definition is invalid; for example $\bar\phi  =\int _{-\pi}^{\pi}d\phi ~\phi f(\phi)$ and subsequently, for the same reason $(\Delta \phi)^2=(<x-<x>)^2>$, will depend up on the choice of the initial value of $\phi$. This means, for example, for a  shift by $\phi_0$, $\phi =\phi' +\phi_0 $, we will find  $\bar\phi \ne\bar\phi' + \phi_0$. The significance of (\ref{a1}) is that this definition is independent of the choice of the initial line.

This leads  us to define, for periodic $\psi(-L/2)=\psi(L/2)$,
$$
V(\gamma )=\int_{-L/2}^{L/2} \psi^*(x+\gamma)x^2\psi(x+\gamma)dx
    $$
\begin{equation}
(\Delta p_x)^2=\int_{-L/2}^{L/2} \psi^*(x)\alpha^2\psi(x)dx,~~\alpha = -i\hbar\partial/\partial_x-<p_x> .
    \label{a4}
\end{equation}
Note that $(\Delta p_x)^2$ is defined in the conventional way. Let us construct
\begin{equation}
(\Delta p_x)^2 V(\gamma )=<\psi_\gamma |\alpha^2\psi_\gamma><\psi_\gamma | x^2\psi_\gamma>,~~\psi_\gamma (x)=\psi(x+\gamma).
    \label{a5}
\end{equation}
In the above both (\ref{a3},\ref{a4}) are independent of the choice of the initial line. Hence, exploiting the Schwartz inequality, we have
\begin{equation}
(\Delta p_x)^2 V(\gamma )\geq |<\psi_\gamma |(x\alpha ) \psi_\gamma >|^2 .
    \label{a6}
\end{equation}
We decompose
\begin{equation}
x\alpha =\frac{1}{2}(x\alpha +\alpha x)+  \frac{1}{2}(x\alpha -\alpha x) ,~~ x\alpha -\alpha x =i\hbar
\label{a7}
\end{equation}
and obtain
\begin{equation}
<\psi_\gamma |(x\alpha +\alpha x ) \psi_\gamma >^*=<\psi_\gamma |(x\alpha +\alpha x ) \psi_\gamma >+2L i\hbar\psi^*(L+\gamma)\psi(L+\gamma ),
    \label{a8}
\end{equation}  

\begin{equation}
Im ~<\psi_\gamma |(x\alpha +\alpha x ) \psi_\gamma >=-L \hbar \psi^*(L+\gamma)\psi(L+\gamma ).
    \label{a9}
\end{equation}
The above steps lead to
\begin{equation}
(\Delta p_x)^2 V(\gamma )\geq \frac{\hbar ^2}{4}(1-L \hbar \psi^*(L+\gamma)\psi(L+\gamma ))^2 .
    \label{a10}
\end{equation}
Renaming $\alpha\equiv x-<x>, \beta\equiv p_x-<p_x>$ we have
\begin{equation}
(\Delta x)^2(\Delta p_x)^2=\int \psi^*\alpha^2\psi dx \int \psi^*\beta^2\psi dx=\int (\alpha^*\psi^*)(\alpha\psi) dx \int (\beta^*\psi^*)(\beta\psi) dx .
    \label{a3}
\end{equation}
Schwartz inequality requires 
\begin{equation}
(\Delta x)^2(\Delta p_x)^2\ge|\int (\alpha^*\psi^*)(\beta \psi) dx|^2=|\int \psi^*(\alpha\beta)\psi dx|^2
    \label{a4}
\end{equation}
In general we have
\begin{equation}
|<f|g>|^2\ge (\frac{<f|g>-<g|f>}{2i})^2.
    \label{a5}
\end{equation}
In the present case with $f\equiv \alpha\psi, g\equiv \beta\psi $ we have
\begin{eqnarray}
&&<f|g>-<g|f>=\int (\psi^*(x+\gamma )x(-i\hbar\partial\psi(x+\gamma))dx -\int ((+i\hbar\partial\psi^*(x+\gamma)x\psi(x+\gamma ))dx \nonumber\\
&&=-i\hbar \int (\psi^*(x+\gamma )x(\partial\psi(x+\gamma))dx 
-i\hbar \int [\partial \{\psi^*(x+\gamma )x\psi(x+\gamma) \}-\psi^*(x+\gamma )\partial(x\psi(x+\gamma))]dx \nonumber\\
&&=-i\hbar \psi^*(x+\gamma )x\psi(x+\gamma) |_0^L +i\hbar \int [-\psi^*(x+\gamma )x\partial\psi(x+\gamma)+\psi^*(x+\gamma )\partial(x\psi(x+\gamma))]dx
    \label{a6}
\end{eqnarray} 
This trivially simplifies to
\begin{equation}
  <f|g>-<g|f>=-i\hbar L  \psi^*(L/2+\gamma )\psi(L/2+\gamma)   +i\hbar \int  \psi^*\psi dx   
            \label{a7}
\end{equation}
for $\psi(L+\gamma)=\psi(\gamma)$, leading to 
\begin{equation}
<f|g>-<g|f>=i\hbar (1-L\psi^*(L+\gamma )\psi(L+\gamma) ).
    \label{a8}
\end{equation}

\begin{equation}
\Delta p_x^2V(\gamma)\ge \frac{\hbar ^2}{4}(1-L\psi^*(L+\gamma )\psi(L+\gamma))^2
    \label{a9}
\end{equation}
The inequality in (\ref{a9}) above applies to all values of $\gamma$. Let us assume that the initial line is at $\gamma =0$ so that, according to our definition $V(0)=\Delta x^2$. 

The following identity will play an important role:
\begin{eqnarray}
&&\frac{dV(y)}{dy}=V'(y)=\frac{d}{dy}[\int^{\beta(y)}_{\alpha(y)}g(x,y)dx]=g(\beta(y),y)\beta'(y)-g(\alpha(y),y)\alpha'(y) \nonumber \\
&&+\int^{\beta(y)}_{\alpha(y)}\frac{\partial g(x,y)}{\partial y}dx .
    \label{b1}
\end{eqnarray}
The above yields the relations, where $f(x)\equiv \psi^*(x)\psi(x)$,
\begin{equation}
V'(\gamma )=-2\int^{L/2}_{-L/2}f(x+\gamma )xdx ,
    \label{b2}
\end{equation}
\begin{equation}
V''(\gamma )=2[1-Lf(L/2+\gamma )]
    \label{b3}
\end{equation}
and the following inequalities,
\begin{equation}
0\leq V(\gamma)\leq L^2/4,~ -L\le V'(\gamma) < L,~ -\infty < V''(\gamma) \leq 2
    \label{b4}
\end{equation}
and finally
\begin{equation}
\bar V=\frac{1}{L}\int ^{L/2}_{-L/2}V(\gamma )d\gamma =L^2/12 .
    \label{b5}
\end{equation}
Following the algebraic argument provided in \cite{j1}, (that can be directly applied to the present case of interest), it can be shown that 
\begin{equation}
\Delta p_x \Delta x\ge \frac{\nu\hbar }{2}(1-12\frac{(\Delta x )^2}{L ^2 }),
    \label{b6}
\end{equation}
where $nu$ is a numerical constant of $\sim 1$. In fact value of $\nu$ for the angular momentum - angle uncertainty relation is derived in \cite{j,j1,j2} but we have not included the corresponding analysis since our main concern was to {\it{demonstrate that the EUP structure can be derived naturally from the assumption of a periodic nature in space in  conventional Quantum Mechanics framework.}} Thus, we claim to have derived the EUP
\begin{equation}
\Delta p_x \Delta x\ge \frac{\hbar }{2}(1-12\frac{(\Delta x )^2}{L ^2 }).
    \label{b7}
\end{equation}

{\bf{The ms. has no associated data.\\
The ms. has no associated code/software.}}


\begin{thebibliography}{99}
 
 \bibitem{gup} M. Maggiore, A Generalized uncertainty principle in quantum gravity, Phys. Lett. B 304,
65 (1993);  R. J. Adler and D. I. Santiago, On gravity and the uncertainty principle, Mod. Phys. Lett.
A 14, 1371 (1999);  F. Scardigli, Generalized uncertainty principle in quantum gravity from micro - black hole
Gedanken experiment, Phys. Lett. B 452, 39 (1999).
\bibitem{rev} Sabine Hossenfelder, Minimal Length Scale Scenarios for Quantum Gravity,  Living Rev.Rel. 16 (2013) 2, ( e-Print: 1203.6191 [gr-qc]); Saurya Das, Elias C. Vagenas, Phenomenological Implications of the Generalized Uncertainty Principle,  Can.J.Phys. 87 (2009) 233-240
 \bibitem{eup}  B. Bolen and M. Cavaglia, (Anti-)de Sitter black hole thermodynamics and the generalized uncertainty principle, Gen. Rel. Grav. 37, 1255 (2005);   C. Bambi and F. R. Urban, Natural extension of the Generalised Uncertainty Principle,
Class. Quant. Grav. 25, 095006 (2008).
 \bibitem{park} M. I. Park, The Generalized Uncertainty Principle in (A)dS Space and the Modification of
Hawking Temperature from the Minimal Length, Phys. Lett. B 659, 698 (2008).
\bibitem{scar}F.Scardigli,
"Hawking temperature for various kinds of black holes from Heisenberg uncertainty principle",
Int.J.Geom.Meth.Mod.Phys. 17 (2020) no.supp01, 2040004 [arXiv:gr-qc/0607010].
 \bibitem{kem} A. Kempf, G. Mangano and R. B. Mann, Hilbert space representation of the minimal length
uncertainty relation, Phys. Rev. D 52, 1108-1118 (1995).
\bibitem{com}H. P. Robertson, The Uncertainty Principle, Phys. Rev. 34, 163 (1929);  E. Schr¨odinger, About Heisenberg uncertainty relation, Bulg. J. Phys. 26, 193 (1999)
[Sitzungsber. Preuss. Akad. Wiss. Berlin (Math. Phys. ) 19, 296 (1930)].
 \bibitem{la}  Matthew J. Lake, Marek Miller, Ray Ganardif  and Tomasz Paterekg, Generalised Uncertainty Relations from
Finite-Accuracy Measurements; Front. Astron. Space Sci., 10, 1087724 (2023), https://doi.org/10.3389/fspas.2023.1087724 .
\bibitem{sg1}Subir Ghosh, Quantum Gravity Effects in Geodesic Motion and Predictions of Equivalence Principle Violation,  Class.Quant.Grav. 31 (2014) 025025 ( e-Print: 1303.1256 [gr-qc]); 
Souvik Pramanik, Subir Ghosh, 
GUP-based and Snyder Non-Commutative Algebras, Relativistic Particle models and Deformed Symmetries and Interaction: A Unified Approach,
Int.J.Mod.Phys.A 28 (2013) 27, 1350131 ( e-Print: 1301.4042 [hep-th]).
\bibitem{s1}F.Scardigli and R.Casadio,
"Gravitational tests of the Generalized Uncertainty Principle",
Eur.Phys.J.C 75 (2015) no.9, 425 [arXiv:1407.0113].
\bibitem{s2} R.Casadio and F.Scardigli,
"Generalized Uncertainty Principle, Classical Mechanics, and General Relativity",
Phys.Lett.B 807 (2020), 135558 [arXiv:2004.04076].
\bibitem{rob} R. Casadio, F. Scardigli, Horizon wave function for single localized particles: GUP and quantum black-hole decay. Eur. Phys. J. C 74, 2685 (2014). https://doi.org/10.1140/epjc/s10052-013-2685-2
 \bibitem{j} D. Judge, On the Uncertainty Relation for $L_z$ and $\varphi$, Phys. Lett. 5 (1963) 189.
\bibitem{j1}D. Judge, Nuovo Cimento 31, 332 (1964)
\bibitem{j2}D. Judge and J.T. Lewis, Phys. Lett. 5, 190(1963).
 \bibitem{p1}G. Busarello, S.  Capozziello,  R. de Ritis,  G.  Longo, 
A.  Rifatto,  C. Rubano, P. Scudellaro,   Apparently periodic Universe, Astronomy and Astrophysics,  283, p. 717 (1994)
 \bibitem{p2}T. J. Broadhurst, R. S. Ellis, D. C. Koo , A. S. Szalay,  Large-scale distribution of galaxies at the Galactic poles, 
 Nature  343, 726–728 (1990)
 \bibitem{p3}A.A. Anselm, Periodic universe and condensate of pseudo-Goldstone field, Physics Letters B,  260,  39-44 (1991)

\bibitem{k}H. Kleinert, Multivalued Fields in Condensed Matter,
Electromagnetism, and Gravitation, (World Scientific,
Singapore, 2008).
\bibitem{jks}P.Jizba, H.Kleinert, F.Scardigli,
"Uncertainty Relation on World Crystal and its Applications to Micro Black Holes",
Phys.Rev.D 81 (2010), 084030 [arXiv:0912.2253].
\end{thebibliography}
\end{document}